\documentclass[a4paper]{jpconf}
\usepackage{graphicx}

\usepackage{amsmath,amssymb}
\usepackage{hyperref}
\usepackage{cite}
\newcommand{\unit}[1]{\ {\rm #1}}

\def\gagamma{g_{a\gamma}}
\def\etal{{\it et al.}}

\begin{document}
\title{Ultralight dark matter searches with KAGRA gravitational wave telescope}

\author{Yuta~Michimura$^{1,2}$, Tomohiro~Fujita$^{3,4}$, Jun'ya~Kume$^4$, Soichiro~Morisaki$^5$, Koji~Nagano$^6$, Hiromasa~Nakatsuka$^7$, Atsushi~Nishizawa$^4$, Ippei~Obata$^8$}
\address{$^1$ Department of Physics, University of Tokyo, Bunkyo, Tokyo 113-0033, Japan}
\address{$^2$ PRESTO, Japan Science and Technology Agency (JST), Kawaguchi, Saitama 332-0012, Japan}
\address{$^3$ Waseda Institute for Advanced Study, Waseda University, Shinjuku, Tokyo 169-8050, Japan}
\address{$^4$ Research Center for the Early Universe, University of Tokyo, Bunkyo, Tokyo 113-0033, Japan}
\address{$^5$ Department of Physics, University of Wisconsin-Milwaukee, Milwaukee, WI 53201, USA}
\address{$^6$ Institute of Space and Astronautical Science, Japan Aerospace Exploration Agency, Sagamihara, Kanagawa 252-5210, Japan}
\address{$^7$ Institute for Cosmic Ray Research, University of Tokyo, Kashiwa, Chiba, 277-8582, Japan}
\address{$^8$ Max-Planck-Institut f{\"u}r Astrophysik, Karl-Schwarzschild-Str. 1, 85741 Garching, Germany}
\ead{michimura@phys.s.u-tokyo.ac.jp}

\begin{abstract}
Among various dark matter candidates, bosonic ultralight fields with masses below 1~eV are well motivated. Recently, a number of novel approaches have been put forward to search for ultralight dark matter candidates using laser interferometers at various scales. Those include our proposals to search for axion-like particles (ALPs) and vector fields with laser interferometric gravitational wave detectors. ALPs can be searched for by measuring the oscillating polarization rotation of laser light. Massive vector fields weakly coupled to the standard model sector can also be searched for by measuring the oscillating forces acting on the suspended mirrors of the interferometers. In this paper, the current status of the activities to search for such ultralight dark matter candidates using a gravitational wave detector in Japan, KAGRA, is reviewed. The analysis of data from KAGRA's observing run in 2020 to search for vector dark matter, and the installation of polarization optics to the arm cavity transmission ports of the interferometer to search for ALPs in future observing runs are underway.
\end{abstract}

\section{Introduction}
For nearly a century after dark matter was first proposed to explain galaxy rotation curves, the existence of dark matter is now regarded as an indispensable ingredient of modern standard cosmology. However, despite numerous experimental and observational efforts to detect dark matter, it has eluded direct detection, and the nature of dark matter still remains a mystery. Conventional dark matter searches have focused on weakly interacting massive particles with masses around $\sim$100~GeV, but so far no particle beyond the standard model has been convincingly detected. There is thus a clear need of searches for alternative candidates, with novel experimental techniques~\cite{NewEraDM}.

Among diverse dark matter candidates spanned over 90 orders of magnitude in mass, bosonic ultralight fields with masses of $10^{-22} \unit{eV} \lesssim m \lesssim 1\unit{eV}$ are cosmologically well-motivated because they behave like non-relativistic fluid in the present universe due to their oscillatory behavior. Recently, a number of novel ideas have been proposed to search for signatures from direct interaction of ultralight dark matter and laser interferometers~\cite{AxionInterferometry,DANCE,ADBC,DenHaixing,NaganoAxion,NaganoAxion2,GroteStadnik,Pierce,MichimuraVector,MorisakiEffect}. Laser interferometers are sensitive to tiny oscillations from such interactions, as clearly demonstrated by the direct detection of gravitational waves with LIGO interferometers in 2015.

Axion-like particles (ALPs) can be searched for by a polarization measurement of laser light~\cite{AxionInterferometry,DANCE,ADBC,DenHaixing,NaganoAxion,NaganoAxion2}. Scalar fields that cause time variation of the fine structure constant or the particle masses can be probed by measuring the optical path length changes in mirrors~\cite{GroteStadnik}. Vector fields weakly coupled to the standard model sector can be searched for by measuring the non-standard oscillating forces on mirrors~\cite{Pierce,MichimuraVector,MorisakiEffect}. Searches for such signals in the actual data from gravitational wave detectors have been recently done, and direct upper limits for scalar field and vector field dark matter were set from GEO 600~\cite{GEOScalar} and LIGO-Virgo~\cite{O1DarkPhoton,O3DarkPhoton}, respectively.

In this paper, we will review our proposals to search for ALPs~\cite{NaganoAxion,NaganoAxion2} and vector bosons~\cite{MichimuraVector,MorisakiEffect} with laser interferometric gravitational wave detectors. The current status of the activities to search for such dark matter candidates using KAGRA detector in Japan is also presented. Throughout this paper, we use the natural unit system, $\hbar=c=\epsilon_0=1$.

\def\miniwid{0.99\hsize}
\begin{figure*}
	\begin{center}
\begin{minipage}[b]{0.49\hsize}
   \begin{center}
   \includegraphics[width=\miniwid]{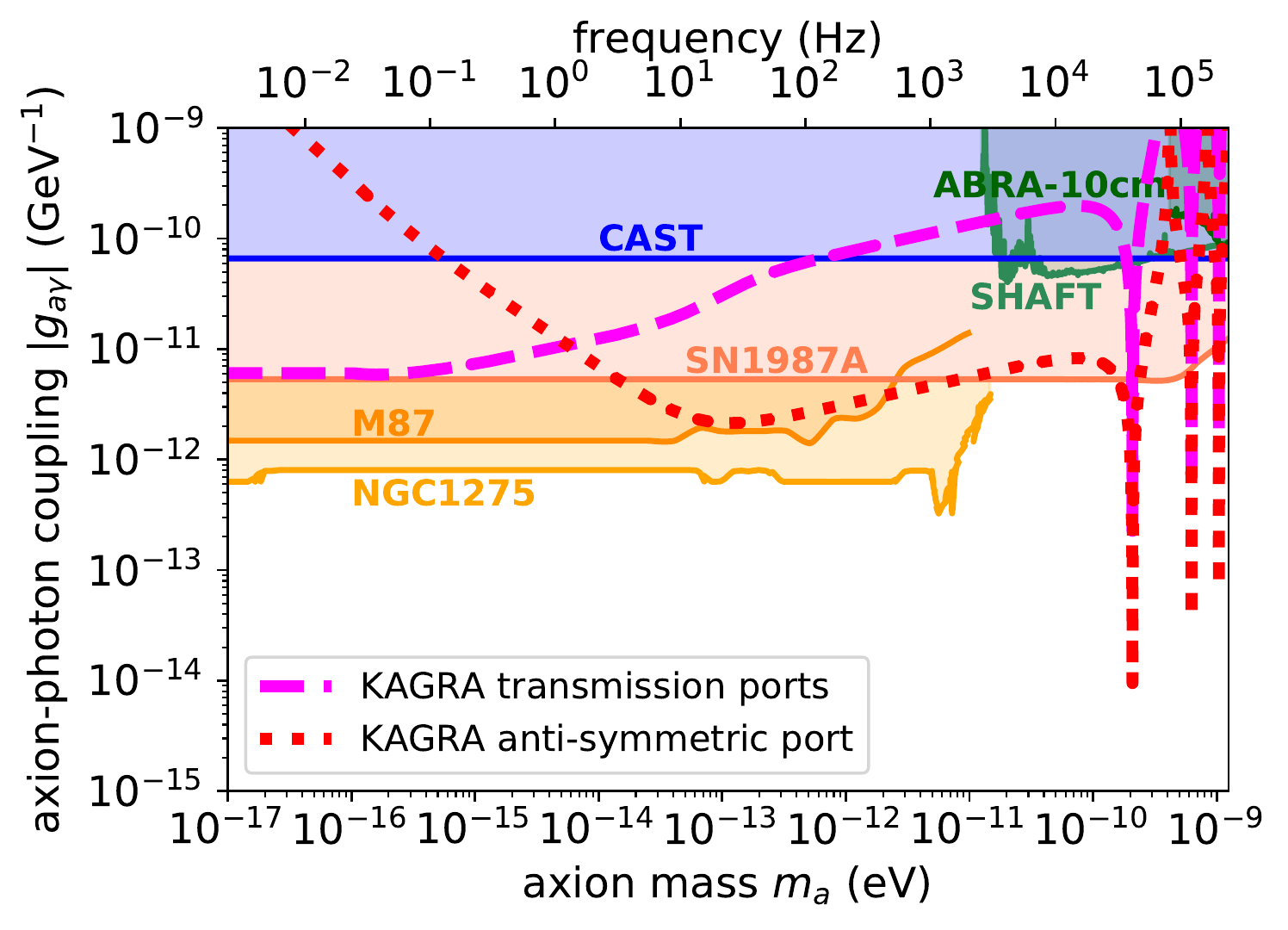} \\
   \end{center}
\end{minipage}   
\begin{minipage}[b]{0.49\hsize}
   \begin{center}
   \includegraphics[width=\miniwid]{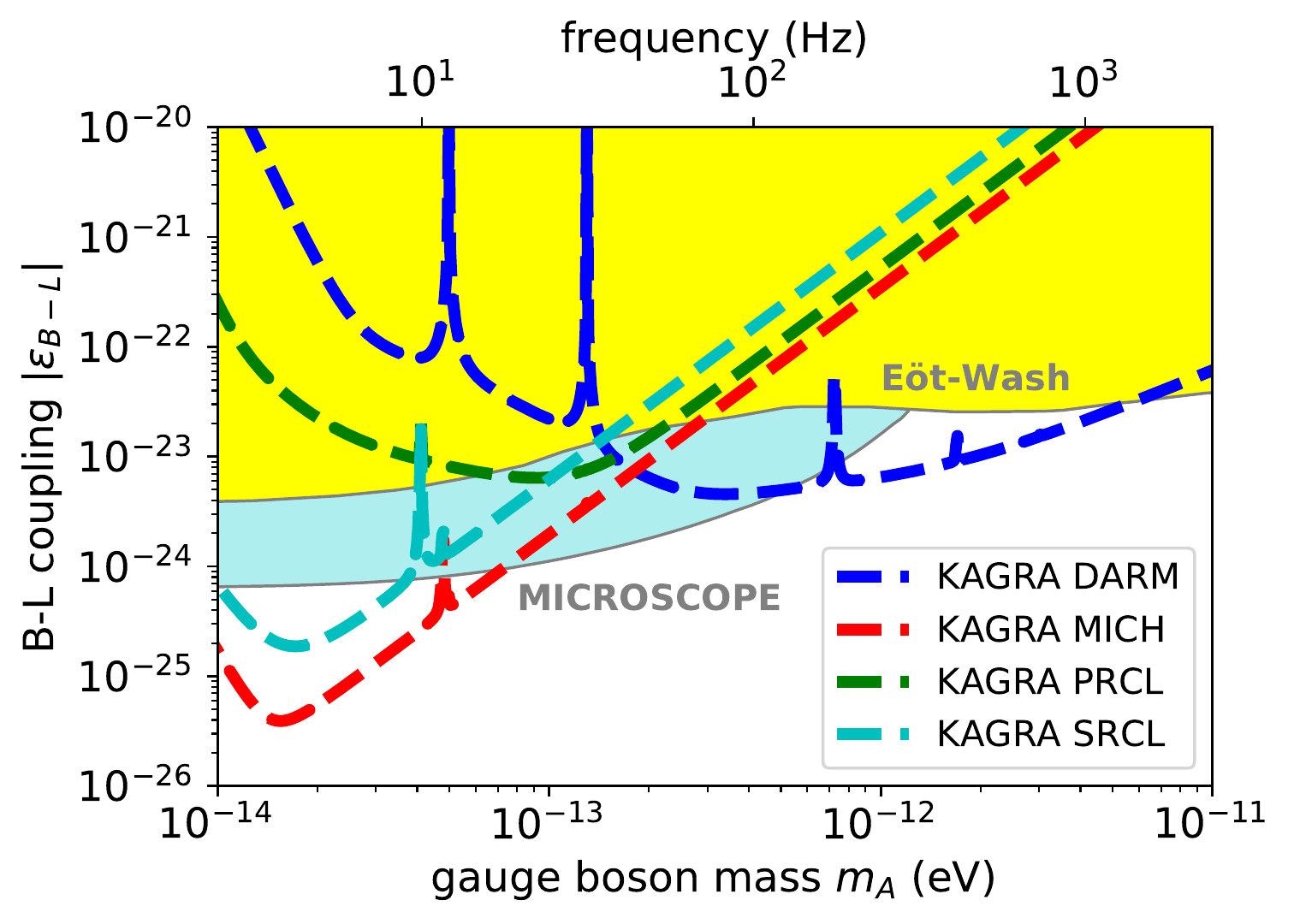} \\
   \end{center}
\end{minipage}
	\caption{\label{Sensitivity} The projected sensitivity for ALPs (left) and $B-L$ vector boson (right) of KAGRA with the measurement time of 1 year. The shaded regions show limits from CAST~\cite{CAST}, SHAFT~\cite{SHAFT}, ABRACADABRA-10cm~\cite{ABRA} experiments, astrophysical bounds from the gamma-ray observations of SN1987A~\cite{SN1987A} and the X-ray observations of M87~\cite{M87} and NGC1275~\cite{NGC1275} galaxies (left), and limits from fifth-fore searches with E{\"o}t-Wash torsion pendulum~\cite{EW2008,EW2012} and MICROSCOPE satellite~\cite{MICROSCOPE2018} (right).}
	\end{center}
\end{figure*}

\section{Axion dark matter search}
Axion is one of the most well known particles beyond the standard model, originally proposed in 1970s to solve the strong CP problem in quantum chromo dynamics (QCD). Besides this ``QCD axion," a multitude of ALPs with the allowed mass range of tens of orders of magnitude are predicted by the string theory. They are also considered to be a leading candidate of dark matter, and there have been enormous efforts to search for their signatures through a variety of experiments and astrophysical observations.

A popular way to search for ALPs utilizes their parity-violating interaction with photons with a coupling constant $\gagamma$. This axion-photon interaction gives a phase velocity difference between left- and right-handed circular polarizations of light. Under the background ALP field $a(t)=a_0 \cos{(m_a t + \delta_\tau(t))}$, the phase velocity difference $\delta c = |c_{\rm L}-c_{\rm R}|=\delta c_0 \sin{(m_a t+\delta_\tau(t))}$ for a wavelength of light $\lambda_{\rm l} = 2 \pi / k_{\rm l}$ is given by
\begin{equation}
 \delta c_0 = \frac{\gagamma a_0 m_a}{k_{\rm l}} \simeq 2.1 \times 10^{-24} \left( \frac{\lambda_{\rm l}}{1064 \unit{nm}} \right) \left( \frac{\gagamma}{10^{-12} \unit{GeV^{-1}}} \right) .
\end{equation}
Here, $m_a$ is the ALP mass and its oscillation frequency is given by $2.4 \unit{Hz} (m_a/(10^{-14} \unit{GeV}))$, and we assumed axion energy density equals local dark matter density, $\rho_a = m_a^2 a_0^2 /2 \simeq 0.4 \unit{GeV/cm^3}$. The phase factor $\delta_\tau(t)$ varies at the coherence timescale of dark matter $\tau=2 \pi / (m_a v^2)$, where $v \simeq 10^{-3}$ being the local velocity of dark matter. This phase velocity difference gives periodic rotation of the plane of linear polarization of light, and this birefringence effect can be searched for with table-top ring cavity experiments such as DANCE~\cite{DANCE}, or with linear cavities in laser interferometric gravitational wave detectors through the ADAM-GD scheme~\cite{NaganoAxion, NaganoAxion2}.

When searching for axion-induced polarization rotation with optical cavities, flipping of the polarization of light upon mirror reflections generally degrades the sensitivity, and the use of wave-plates~\cite{AxionInterferometry} or a bow-tie cavity~\cite{DANCE,ADBC}, for example, is necessary to overcome this difficulty. However, when the finite light-travelling time between mirrors is not negligible compared with the axion oscillation period, linear cavities can also be sensitive. Especially, when the round-trip time of light equals odd-multiples of the axion oscillation period, the sensitivity is enhanced by multiple number of trips inside the cavity.

Long-baseline linear cavities in the arms of gravitational wave detectors are suitable to take advantage of this effect, and the ALP search can be done by monitoring the polarization state of laser light at different interferometer ports through ADAM-GD scheme. The ALP search can be done by adding a few optics and without sacrificing the sensitivity to gravitational waves, since signatures of gravitational waves are imprinted in the phase of the carrier polarization, while ALP signal is in the polarization orthogonal to the input beam.

\begin{figure}[t]
  \begin{center}
    \includegraphics[width=0.82\hsize]{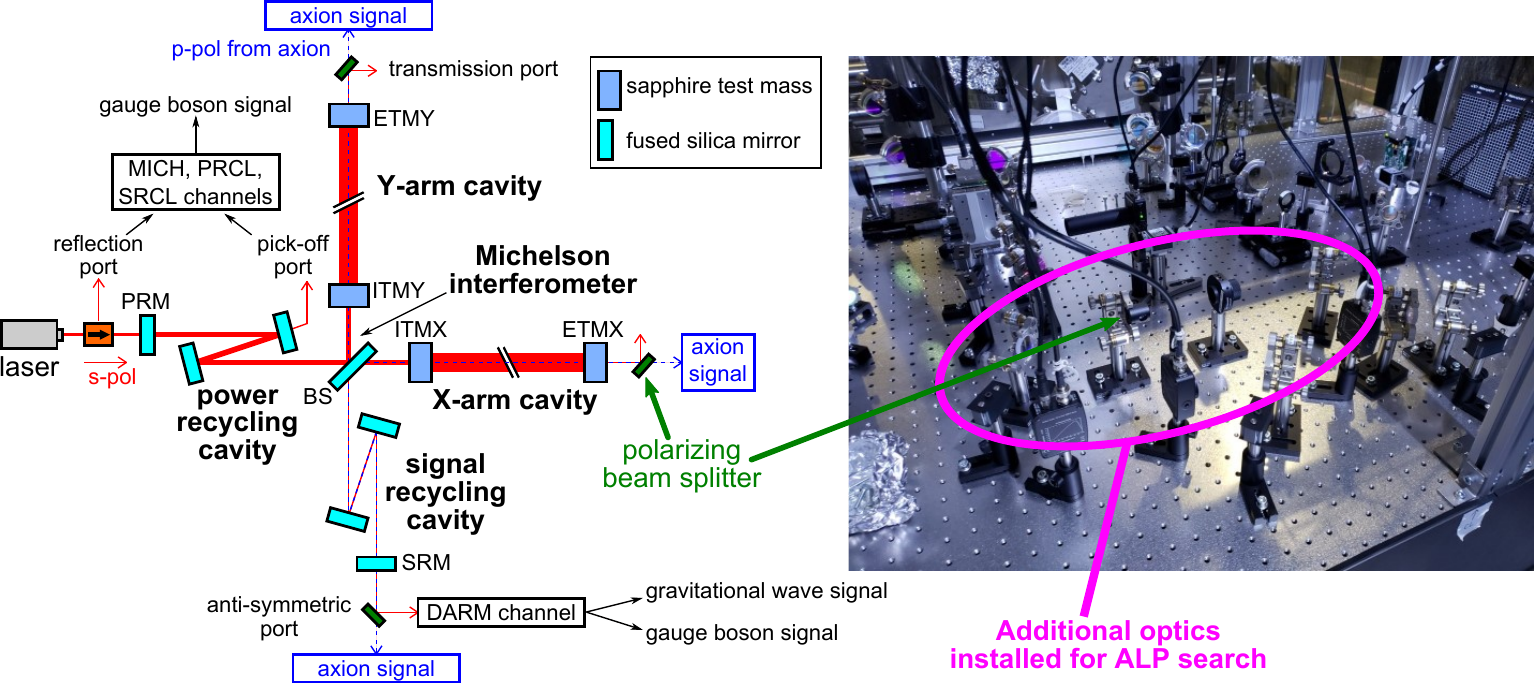}
    \caption{\label{config} The schematic of KAGRA for ALP and vector dark matter searches (left) and the photo of polarization optics installed at the X-arm cavity transmission for ALP search (right).}
  \end{center}
\end{figure}

Figure~\ref{Sensitivity} (left) shows the projected sensitivity of the KAGRA detector, computed with a method described in Refs.~\cite{NaganoAxion, NaganoAxion2}. Since the arm cavities are overcoupled, the search using the cavity reflected beams at the anti-symmetric port has generally better sensitivity~\cite{NaganoAxion}, but the search using the arm cavity transmitted beams has better sensitivity at lower masses of ALPs (Fig.~\ref{config}). This is because the light at the transmission ports travels in the cavity for odd-number of times, and the polarization rotation effect is not cancelled out by polarization flipping upon mirror reflections~\cite{NaganoAxion2}. In both cases, the sensitivity peaks at $m_a=\pi (2N-1)/L_{\rm cav}$, where $L_{\rm cav}=3$~km is the arm cavity length and $N \in \mathbb{N}$. At the time of writing, installation of the polarization optics to the X-arm cavity transmission is completed, and those for Y-arm will be installed soon to prepare for the O4 observing run, which is planned to start in 2022.

\section{Vector dark matter search}
Another promising candidate of dark matter is a massive vector field arising as a gauge boson of $U(1)_B$ or $U(1)_{B-L}$ gauge symmetry, where $B$ and $L$ are the baryon and lepton numbers, respectively. The vector dark matter gives a non-standard oscillating force to the test masses through its coupling with baryons and/or leptons. Recently, ideas to use gravitational wave detectors to search for such signatures have been proposed, and the constraints better than those from equivalence principle tests~\cite{EW2008,EW2012,MICROSCOPE2018} were obtained from the actual search using the LIGO-Virgo data for the $U(1)_B$ case~\cite{O3DarkPhoton}.

Under the background vector field $\vec{A}(t,\vec{x})=A_0 \vec{e}_A \cos{(m_A t - \vec{k} \cdot \vec{x} + \delta_\tau(t))}$ at location $\vec{x}$, the oscillating force $\vec{F}=F_0 \vec{e}_A \sin{(m_A t - \vec{k} \cdot \vec{x} + \delta_\tau(t))}$ on an object with a mass of $M$ and a charge of $q_D$ is given by
\begin{equation}
 F_0 = \epsilon_D e q_D m_A A_0 \simeq \epsilon_D q_D \times 6.1 \times 10^{-16} \unit{N} \simeq \epsilon_D M \times 3.6 \times 10^{11} \unit{m/s^2} \times
    \begin{cases}
        \sim1   &   \text{(for $D=B$)}  \\
        \sim0.5 &   \text{(for $D=B-L$)}
    \end{cases} .
\end{equation}
Here, $\epsilon_D$ is the gauge coupling constant of $U(1)_D$ $(D=B$ or $B-L)$ normalized to the electromagnetic coupling, $\vec{e}_A$ is the unit vector parallel to $\vec{A}$, and $k=m_A v$. The amplitude of the oscillating acceleration from this force is proportional to $q_D/M$, and this differs between materials. For a $U(1)_B$ gauge boson, $q_B/M \simeq 1/m_{\rm n}$ is almost identical between different materials, where $m_{\rm n}$ is the neutron mass. Whereas for a $U(1)_{B-L}$ gauge boson, $q_{B-L}/M \simeq 0.5/m_{\rm n}$ is more distinguishable between materials which have different neutron ratios.

Gravitational wave detectors are highly sensitive to measure differential arm length (DARM) changes in two perpendicular arm cavities, since gravitational waves create DARM changes. However, accelerations from the vector field are mostly common to four arm-cavity test mass mirrors made of the same material, and most of the vector dark matter signal is canceled out. The sensitivity using the DARM channel comes from the residual effect due to the oscillation phase difference between locations, and the effect from the finite light-traveling time. In the case of long-baseline interferometers, the latter is significant since the test masses oscillate while the light is traveling in the arm in the mass range which satisfies $v \ll m_A L$~\cite{MorisakiEffect}.

In the case of KAGRA, auxiliary length channels such as differential Michelson interferometer length (MICH), power recycling cavity length (PRCL) and signal recycling cavity length (SRCL) can be also used, as auxiliary parts of the interferometer consist of sapphire test masses and fused silica auxiliary mirrors (Fig.~\ref{config}(right)). For sapphire and fused silica, $q_{B-L}/M$ is 0.51 and 0.501, respectively, and the search using the auxiliary channels can be more sensitive than the main DARM channel~\cite{MichimuraVector}. This is peculiar to KAGRA, as other gravitational wave detectors such as LIGO and Virgo use fused silica for all the mirrors.

Figure~\ref{Sensitivity} (right) shows the projected sensitivity of KAGRA for $U(1)_{B-L}$ gauge boson, computed using a method described in Refs.~\cite{MichimuraVector,MorisakiEffect}. At lower masses, the auxiliary channels have better sensitivity than the DARM channel, since they can take advantage of different $q_{B-L}/M$ for sapphire and fused silica, while the effect is suppressed by a factor of $\sim v$ in the DARM at lower masses. At higher masses, the DARM channel has better sensitivity since the effect of finite light-traveling time becomes significant and the displacement sensitivity of auxiliary channels are much worse than the DARM.

The signature from the vector dark matter in the data is similar to that from ALPs, and is nearly a monochromatic signal with a frequency linewidth of $\sim m_A/10^6$. Therefore, a Fourier analysis provides an efficient way to search for such signals. At the time of writing, we are working on the analysis of data from KAGRA's first observing run O3GK in 2020~\cite{KAGRA}. The calibrations of PRCL and MICH signals were done after the observing run, since they were not originally considered to be used as science data and were not calibrated carefully. The basic pipeline has been developed, and the spectral line noise veto such as veto using the linewidth of the signal and consistency check between the segments is underway.

\section{Summary and outlook}
Laser interferometric gravitational wave detectors open up new possibilities for ultralight dark matter search. ALPs search can be done by searching for oscillating polarization rotation of laser light at different interferometer ports. A massive vector field as a gauge boson of $U(1)_B$ or $U(1)_{B-L}$ can also be searched for by measuring oscillating length changes. These signals are nearly monochromatic, and can be efficiently searched for with a Fourier analysis. Within the KAGRA project, the installation of the polarization optics to search for ALPs in the next observing run O4 in 2022 is underway. The analysis of data from KAGRA's observing run in 2020 to look for signatures from vector dark matter is also underway. KAGRA is a unique detector which can provide auxiliary length channels to search for differential displacement of sapphire test masses and fused silica auxiliary mirrors, which have different $B-L$ charge densities. KAGRA with improved sensitivity in the future observing runs will play an important role not only in the global network of gravitational wave observations, but also in our quest to reveal the nature of dark matter.

\ack
This work was supported by JSPS KAKENHI Grant Nos. 18H01224, 18K13537, 19H01894, 19J21974, 20H04726, 20H05850, 20H05854, 20H05859, 20J01928, 20J218661, JST PRESTO Grant No. JPMJPR200B, and NSF PHY-1912649. K.N., H.N, A.N. and I.O acknowledge the support from JSPS Research Fellowship, the Advanced Leading Graduate Course for Photon Science, Research Grants from Inamori Foundation, and JSPS Overseas Research Fellowship, respectively.


\section*{References}
\begin{thebibliography}{99}
\bibitem{NewEraDM}
Bertone~G, Tait~T~M~P 2018 {\em Nature\/} \href{https://doi.org/10.1038/s41586-018-0542-z}{{\bf 562}, 51}

\bibitem{AxionInterferometry}
DeRocco~W, Hook~A 2018 {\em Phys. Rev. D} \href{https://doi.org/10.1103/PhysRevD.98.035021}{{\bf 98}, 035021}

\bibitem{DANCE}
Obata~I, Fujita~T, Michimura~Y 2018 {\em Phys. Rev. Lett.\/} \href{https://doi.org/10.1103/PhysRevLett.121.161301}{{\bf 121}, 161301}

\bibitem{ADBC}
Liu~H, Elwood~B~D, Evans~M, Thaler~J 2019 {\em Phys. Rev. D\/} \href{https://doi.org/10.1103/PhysRevD.100.023548}{{\bf 100}, 023548}

\bibitem{DenHaixing}
Martynov~D,  Miao~H 2020 {\em Phys. Rev. D\/} \href{https://doi.org/10.1103/PhysRevD.101.095034}{{\bf 101}, 095034}

\bibitem{NaganoAxion}
Nagano~K, Fujita~T, Michimura~Y, Obata~I 2019 {\em Phys. Rev. Lett.\/} \href{https://doi.org/10.1103/PhysRevLett.123.111301}{{\bf 123}, 111301}

\bibitem{NaganoAxion2}
Nagano~K, Nakatsuka~K, Morisaki~K, Fujita~T, Michimura~Y, Obata~I 2021 {\em Phys. Rev. D\/} \href{https://doi.org/10.1103/physrevd.104.062008}{{\bf 104}, 062008}

\bibitem{GroteStadnik}
Grote~H, Stadnik~Y~V 2019, {\em Phys. Rev. Research\/} \href{https://doi.org/10.1103/PhysRevResearch.1.033187}{{\bf 1}, 033187}

\bibitem{Pierce}
Pierce~A, Riles~K, Zhao~Y 2018 {\em Phys. Rev. Lett.\/} \href{https://doi.org/10.1103/PhysRevLett.121.061102}{{\bf 121}, 061102}

\bibitem{MichimuraVector}
Michimura~Y, Fujita~Y, Morisaki~S, Nakatsuka~H, Obata~I 2020 {\em Phys. Rev. D\/} \href{https://doi.org/10.1103/physrevd.102.102001}{{\bf 102}, 102001}

\bibitem{MorisakiEffect}
Morisaki~S, Fujita~T, Michimura~Y, Nakatsuka~H, Obata~I 2021 {\em Phys. Rev. D\/} \href{https://doi.org/10.1103/physrevd.103.l051702}{{\bf 103}, L051702}

\bibitem{GEOScalar}
Vermeulen~S~M \etal, \href{https://arxiv.org/abs/2103.03783}{arXiv:2103.03783}

\bibitem{O1DarkPhoton}
Guo~H-K, Riles~K, Yang~F-W, Zhao~Y 2019 {\em Commun. Phys.\/} \href{https://doi.org/10.1038/s42005-019-0255-0}{{\bf 2}, 155}

\bibitem{O3DarkPhoton}
The LIGO Scientific Collaboration, the Virgo Collaboration, the KAGRA Collaboration, \href{https://arxiv.org/abs/2105.13085}{arXiv:2105.13085}

\bibitem{CAST}
CAST Collaboration 2017 {\em Nat. Phys.\/} \href{https://doi.org/10.1038/nphys4109}{{\bf 13}, 584}

\bibitem{SHAFT}
Gramolin~V~A, Aybas~D, Johnson~D, Adam~J, Sushkov~A~O 2021 {\em Nat. Phys.\/} \href{https://doi.org/10.1038/s41567-020-1006-6}{{\bf 17}, 79}

\bibitem{ABRA}
Salemi~C~P, Foster~J~W, Ouellet~J~L \etal\ 2021 {\em Phys. Rev. Lett.\/} \href{https://doi.org/10.1103/PhysRevLett.127.081801}{{\bf 127}, 081801}

\bibitem{SN1987A}
Payez~A, Evoli~C, Fischer~T, Giannotti~M, Mirizzi~A, Ringwald~A 2015 {\em J. Cosmol. Astropart. Phys.\/} \href{https://doi.org/10.1088/1475-7516/2015/02/006}{{\bf 02}, 006}

\bibitem{M87}
Marsh~M~C~D \etal 2017 {\em J. Cosmol. Astropart. Phys.\/} \href{https://doi.org/10.1088/1475-7516/2017/12/036}{{\bf 12}, 036}

\bibitem{NGC1275}
Reynolds~C~S \etal 2020 {\em Astrophys. J.\/} \href{https://doi.org/10.3847/1538-4357/ab6a0c}{{\bf 890}, 59}

\bibitem{EW2008}
Schlamminger~S, Choi~K-Y, Wagner~T~A, Gundlach~J~H, Adelberger~E~G 2008 {\em Phys. Rev. Lett.\/} \href{https://doi.org/10.1103/PhysRevLett.100.041101}{{\bf 100}, 041101}

\bibitem{EW2012}
Wagner~T~A, Schlamminger~S, Gundlach~J~H, Adelberger~E~G 2008 {\em Classical Quantum Gravity\/} \href{https://doi.org/10.1088/0264-9381/29/18/184002}{{\bf 29}, 184002}

\bibitem{MICROSCOPE2018}
Berg{\'e}~J, Brax~P, M{\'e}tris~G, Pernot-Borr{\`a}s~M, Touboul~P, Uzan~J-P 2018 {\em Phys. Rev. Lett.\/} \href{https://doi.org/10.1103/PhysRevLett.120.141101}{{\bf 120}, 141101}

\bibitem{KAGRA}
KAGRA Collaboration 2020 {\em Prog. Theor. Exp. Phys.} \href{https://doi.org/10.1093/ptep/ptaa125}{{\bf 2020}, 05A101}

\end{thebibliography}
\end{document}